\documentclass[aps,twoside,prb,twocolumn,showpacs,showkeys,citeautoscript]{revtex4-1}
\usepackage{graphicx}
\usepackage{amsmath,amssymb,amstext}

\newcommand{\dd}{d}
\newcommand{\bvec}[1]{\boldsymbol{#1}}

\begin{document}

\title{Thermoelectric Effect of Correlated Metals - Band Structure Effects and the Breakdown of Mott's Formula}

\author{Jonathan M. \surname{Buhmann}}
\affiliation{Institute for Theoretical Physics, ETH Z\"{u}rich, 8093 Z\"{u}rich, Switzerland}
\author{Manfred \surname{Sigrist}}
\affiliation{Institute for Theoretical Physics, ETH Z\"{u}rich, 8093 Z\"{u}rich, Switzerland}

\begin{abstract}
     We study the thermoelectric effect of two-dimensional metals on a square lattice within semi-classical Boltzmann transport theory with particular focus on electron-electron scattering. We compute the electrical conductivity and the Seebeck coefficient as a function of band filling and temperature for generically chosen hopping parameters in a two-dimensional tight binding model. The Boltzmann equation is solved numerically after computing the full collision integral taking the angular and radial degrees of freedom into account. These degrees of freedom of the collision integral, neglected in the standard single-relaxation-time approximation, play an important role if the transport coefficients show unconventional features. Within our detailed numerical simulation, we show that the widely used Mott formula to compute the Seebeck effect is not sufficient to describe the thermoelectric effect in the presence of strong electron-electron scattering. Furthermore, we study the Seebeck coefficient and its temperature dependence in the vicinity of a Lifshitz transition and demonstrate that it shows remarkable parallels to transport features near a quantum critical point.
     \begin{center}
     This paper is published as ''Phys. Rev. B \textbf{88}, 115128 (2013)''\\
     \url{http://link.aps.org/doi/10.1103/PhysRevB.88.115128}
     \end{center}
\end{abstract}


\maketitle

\section{Introduction}

The search for an efficient conversion of heat, in many cases waste heat, into electrical energy \cite{rowe1978,rowe1994} is the motivation for a long-standing investigation of thermoelectricity of various materials\cite{pei_band_2012,lofwander22007,schweitzer1991,heremans2004,koshibae2000,narayan_unconventional_2012}. While for technical applications semiconductors are so far the materials with the highest figures of merit or efficiency, there has been also much effort to examine conditions under which metals would perform optimally. Metals are typically good conductors but have usually a rather low Seebeck coefficient $Q$ and figure of merit.

Besides the interest in potential applications, the study of the thermoelectric effect in metals is also of more conceptual interest. In recent years, the study of quantum critical phenomena has received great attention\cite{pourret_magnetic_2013,pfau_thermoelectric_2012,pfau_interplay_2013,izawa_thermoelectric_2010,kim_thermopower_2010,machida_thermoelectric_2012}. The measurement of the Seebeck coefficient is a useful tool to detect deviations from Fermi liquid behavior as, for example, found in the vicinity of quantum phase transitions. In order to use thermopower as an identifying criterion, one first has to fully understand the phenomenology in the non-critical region of the phase diagram. In a previous study\cite{buhmannLSCO} on the normal-state transport of overdoped La$_{2-x}$Sr$_{x}$CuO$_{4}$, we have demonstrated that one can find highly unconventional behavior of the Seebeck coefficient in a model of a correlated metal with a nontrivial band structure.

In our present paper we address the thermoelectric properties of interacting two-dimensional Fermi liquids on a square lattice. This choice is motivated in parts by the wide spread studies of layered transition-metal-oxide systems\cite{sagarna_electronic_2012,ruegg_role_2009_book,ruegg_aspects_2008,ohta_giant_2007,bocher_perovskite_2008,moser_influence_2011}. Moreover,  it was shown by Hicks and Dresselhaus that for metallic systems low-dimensionality can have a beneficial impact on thermopower\cite{hicks}. We discuss the influence of momentum space constraints for two-particle scattering on the charge transport in general. These constraints are influenced essentially by the Fermi surface geometry. In particular, we address two essential features in this context, the onset of umklapp scattering upon varying electron band fillings and the Lifshitz transition between an electron like and a hole like Fermi surface. Both can be tuned by changing the chemical potential and allow for tuning of the thermo-electricity, in principle. 

Our study is based on the numerical solution of the Boltzmann transport equation for a two-dimensional electronic system with a single-orbital tight-binding band structure. We concentrate on the effect of impurity and especially electron-electron scattering, where the latter contributes intriguing features to the transport properties. In order to highlight the contribution of electron-electron scattering, we completely neglect electron-phonon scattering in this study. For simple metals, it is well known that phonons play an important role regarding the transport of charge and heat for temperatures on the order of 100\,K. In the context of oxides or other strongly correlated materials, the contribution from electron phonon scattering is often less clear. By neglecting phonons, we do not aim for a realistic description of any real system, but we point out the nontrivial effects that electron-electron-scattering can have.

We start this paper with a short introduction of our model followed by a discussion of the structure of the electric conductivity as a function of the band filling. In Sec. \ref{sec:Thermopower and the breakdown of Motts formula} we analyze the validity of the widely used Mott's formula in the context of interacting Fermi liquids. We compare predictions of Mott's formula to an explicit calculation of the Seebeck coefficient based on our computational solution of the Boltzmann equation. Given the Seebeck coefficient as a function of the temperature, we discuss unconventional temperature dependencies of the Seebeck coefficient in the vicinity of a van Hove singularity and the onset of umklapp scattering in Sec. \ref{sec:Unconventional thermopower related to the Lifshitz transition and the umklapp edges}. Eventually we demonstrate that non-Fermi-liquid behaviors, similar to those observed in the vicinity of a quantum critical point (QCP), can also naturally occur in interacting Fermi liquids without any loss of quasiparticle properties. 

\section{A semiclassical model of charge and heat transport}

\subsection{The tight-binding dispersion}

We consider a system of electrons on a two-dimensional square lattice. This system is represented by a quadratic Brillouin zone in reciprocal space. The dispersion of the quasiparticles is conveniently described with a two-dimensional single-orbital tight-binding model, where we restrict ourselves for simplicity to nearest- and next-to-nearest-neighbor hopping,
\begin{align}
     \varepsilon_{\bvec{k}} & = -2t(\cos k_{x}+ \cos k_{y}) + 4t' \cos k_{x}\cos k_{y}
.
\label{eq: tight binding model}
\end{align}
 
It is useful to include next-nearest-neighbor hopping $t'$ to avoid particle-hole symmetry in the model, which also introduces flexibility for
the Fermi-surface curvature at intermediate filling levels.
The nearest-neighbor hopping integral $t$ defines the natural energy scale of our model. It is related to the bandwidth via $w=8\,t$, as long as $t'<0.5\,t$. Tuning the chemical potential $\mu$ changes the band filling and, thus, allows for a scan through the entire band from the bottom to the top.

\subsection{Scattering mechanisms and momentum relaxation}
\label{subsec: scattering mechanisms}

While impurity scattering gives rise to a rather obvious contribution to the momentum relaxation, two-particle scattering is more complicated in this respect.
The electrical conductivity of ultra-clean materials in the low-temperature regime is actually limited by the umklapp-scattering processes, the only possibility to relax the momentum of the Fermi sea to the lattice, in the absence of impurity and electron-phonon scattering. Without momentum relaxation no resistive state can be realized. The presence and extent of umklapp scattering, however, depends on the shape of the Fermi surface as well as the Brillouin zone geometry. This opens the possibility for finding rich features in the conductivity and the thermoelectric power by changing the electronic dispersion parameters.

\begin{figure*}[t]
\begin{center}

\includegraphics[width=0.9\textwidth]{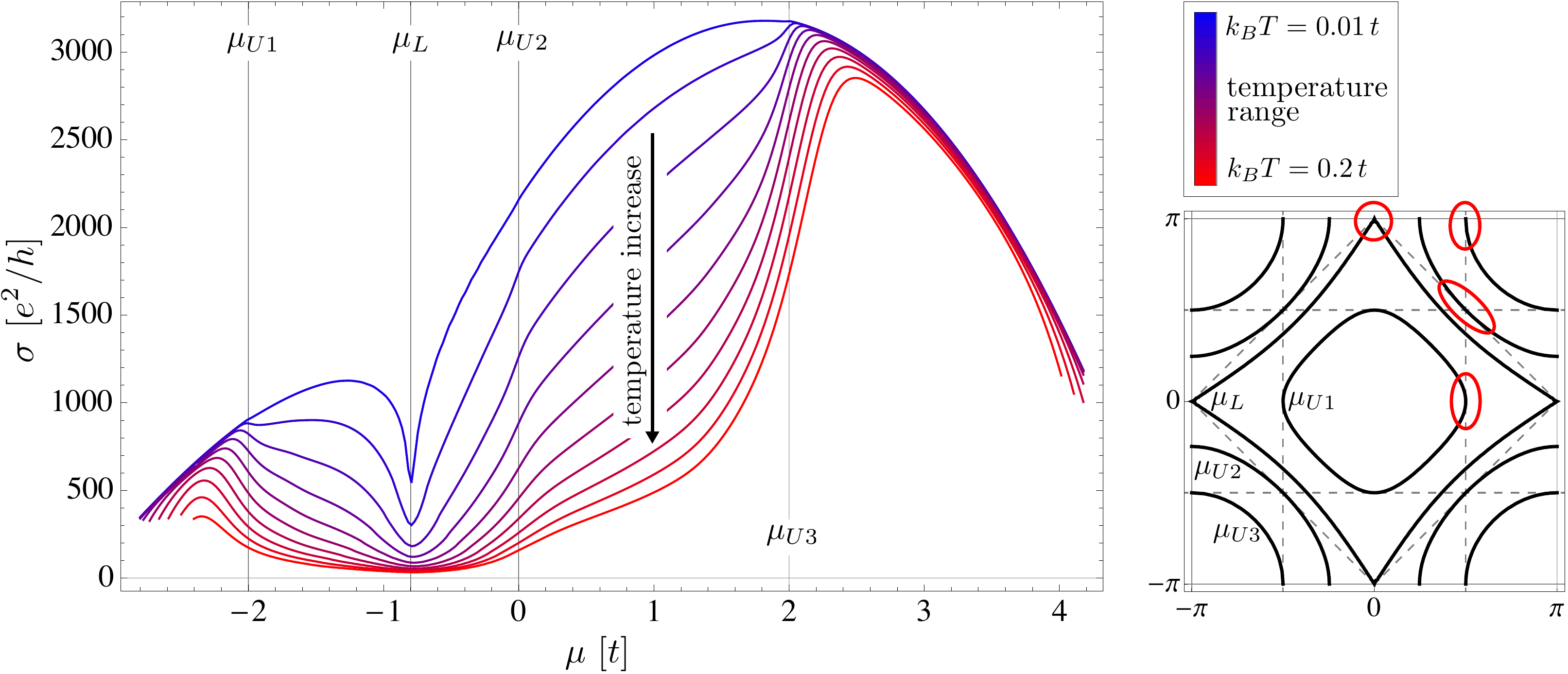}
\caption{Conductivity as a function of the band filling labeled by the chemical potential $\mu$. The conductivity for a two-dimensional electron system on a square lattice with a dispersion \eqref{eq: tight binding model} for the generic choice of $t'=0.2\,t$ is plotted. To the right we show the Fermi surface for the special filling levels, indicated by the horizontal lines in the conductivity plot. The red circles highlight regions of the Fermi surface that touche a geometric umklapp-scattering boundary.}
\label{fig: conductivityBand}

\end{center}
\end{figure*}

As an illustrative example let us consider a circular Fermi surface around the $ \Gamma $ point with a diameter smaller than $\pi/a$ (half the length of a basic reciprocal lattice vector). In this case no umklapp-scattering processes for particles on the Fermi surface are possible. Such umklapp is only
possible on Fermi surfaces which cross the umklapp lines (dashed lines within the Brillouin zone in the right panel of Fig. \ref{fig: conductivityBand}). In this case, the maximal momentum transfer within a scattering event is sufficient for one particle to be scattered to the Fermi surface in a neighboring Brillouin zone. Thus, two-particle umklapp scattering processes which conserve both energy and the lattice momentum are possible, and they transfer momentum to the lattice.

At finite temperature, states involved in scattering are thermally smeared around the Fermi surface. This yields finite but exponentially small scattering contributions even for a Fermi surface just below the umklapp threshold, leading to $ \rho \propto \exp\{- T_0/T\}$ with $ k_BT_0 $. The resistivity can be interpreted as a measure for the difference between the chemical potential and the band filling level, allowing for umklapp scattering at zero temperature. Note that if just a small segment of the Fermi surface crosses the umklapp lines, only states in the close vicinity to the crossing points contribute to this scattering, which is consequently highly anisotropic. Only with this umklapp contribution can the well-known Fermi liquid $ T^2 $ scaling of the resistivity at low temperature be observed.

In our simulation, we study this system over a temperature range from $k_{B}T=0.01\,t$ to $k_{B}T=0.2\,t$, measured in units of the nearest-neighbor hopping energy. For typical metals, this corresponds to a temperature range from $\mathcal O(10\,\textrm{K})$ to a few hundred Kelvin. The relative strength of the impurity scattering potential compared to the electron-electron scattering is chosen in such a way that impurity scattering dominates at the lowest temperatures of our study, while two-particle scattering strongly dominates at high temperature. This crossover from isotropic to anisotropic scattering has important consequences that are discussed below.

\subsection{Seebeck coefficient within Boltzmann transport theory}

The Seebeck coefficient $Q$ is defined by the electric field $\bvec{\mathcal E}$ that is generated in the presence of a thermal gradient for vanishing electric current (open circuit geometry),
\begin{align}
     \bvec{\mathcal E}=-Q\,\bvec{\nabla}T.
\label{eq:def of Q}
\end{align}
The mechanism that drives the thermoelectric effect includes thermal diffusion of charge carriers and drag effects. Here we ignore the latter and concentrate on purely electronic contributions.

For a metal in the low-temperature limit, $T \ll T_F $, the coefficient $Q$ is often computed by taking the logarithmic derivative of the electrical conductivity with respect to the chemical potential \cite{goldsmid},
\begin{align}
     Q_{\text{Mott}}=-\frac{\pi^{2}}{3}\frac{k_{B}^{2}T}{e} \frac{d}{d\mu} \ln \sigma (\mu).
\label{eq: Mott}
\end{align}
This expression, known as Mott's formula, is based on a Sommerfeld expansion of the integral kernels in the general transport coefficients\cite{ashcroftmermin}. An important condition for this approximation is scattering behavior that is isotropic over all the Fermi surface, a property which we discuss in detail in Sec. \ref{sec:Thermopower and the breakdown of Motts formula}.

For the computation of the Seebeck coefficient beyond Mott's formula, we make use of Boltzmann transport theory. The solution of the Boltzmann equation in the presence of an external thermal gradient and an induced electric field,
\begin{align}
     -\frac{\partial f(\bvec{k})}{\partial \varepsilon_{\bvec{k}}}\bvec{v}_{\bvec{k}}\cdot\left(\bvec{\nabla}_{\bvec{r}}T\frac{\varepsilon_{\bvec{k}}-\mu}{T} - \bvec{\mathcal E}\right)=\left.\frac{\partial f(\bvec{k})}{\partial t}\right|_\text{coll},
\label{eq: boltzmann equation with thermal gradient}
\end{align}
is the non-equilibrium steady-state distribution function $f=f_{0}+\delta f$, which can be used to compute electrical and thermal currents.

The collision integral on the right-hand side is a weighted integral over all possible collision processes that the charge carriers can undergo. We consider two contributions to the collision integral, impurity potential scattering and two-particle scattering. While the collision integral from impurity scattering is relatively simple with only energy conserving scattering rates $\Gamma^{\text{imp}}$, the collision integral for two-particle scattering processes consists of a complicated multi-dimensional integration over all momenta with a strongly peaked kernel due to the energy and momentum conservation in the scattering transition rates $\Gamma^{ee}$. The transition rates are computed via Fermi's golden rule from an repulsive on-site Hubbard-$U$-type coupling ($\Gamma^{ee}$) and from nonmagnetic impurity scattering modeled by isotropic delta-potential scatterers ($\Gamma^{\text{imp}}$),

\begin{align}
     \left.\frac{\partial f(\bvec{k}_{1})}{\partial t}\right|_\text{coll}&=-\int\frac{d \bvec{k}_{2}}{(2\pi)^{2}}\frac{\dd \bvec{k}_{3}}{(2\pi)^{2}}\frac{d \bvec{k}_{4}}{(2\pi)^{2}}\Gamma^{ee}(\bvec{k}_{1},\bvec{k}_{2},\bvec{k}_{3},\bvec{k}_{4})\nonumber\\&\phantom{=}\ \times\bigl\{f(\bvec{k}_{1})f(\bvec{k}_{2})(1-f(\bvec{k}_{3}))(1-f(\bvec{k}_{4}))\nonumber\\
     &\phantom{=}\ -(1-f(\bvec{k}_{1}))(1-f(\bvec{k}_{2}))f(\bvec{k}_{3})f(\bvec{k}_{4})\bigr\}\nonumber\\
     &\phantom{=}\ -\int\frac{d \bvec{k}_{2}}{(2\pi)^{2}}\Gamma^{\text{imp}}(\bvec{k}_{1},\bvec{k}_{2})\bigl\{f(\bvec{k}_{1})(1-f(\bvec{k}_{2}))\nonumber\\&\phantom{=}\ -(1-f(\bvec{k}_{1}))f(\bvec{k}_{2})\bigr\}.\label{eq:collisionIntegral}
\end{align}

In the spirit of linear response theory, we first linearize the Boltzmann equation \eqref{eq: boltzmann equation with thermal gradient} in the external fields. We solve the linearized Boltzmann equation by computing the collision integral \eqref{eq:collisionIntegral} in a discretized Brillouin zone. The discretization scheme is designed to match the curvature of the Fermi surface, allowing for a high resolution analysis of the non-equilibrium distribution function in the low-energy sector of the momentum space. With this numerical approach, introduced in Ref. \onlinecite{buhmannLSCO}, the anisotropy and the full energy dependence of two-particle scattering events are explicitly taken into account. Computing the full collision integral represents a substantial sophistication compared to the conventional treatment of the Boltzmann equation within the single-relaxation-time approximation and is essential to derive the results below.

The distribution function $f$ that solves the Boltzmann equation \eqref{eq: boltzmann equation with thermal gradient} in the presence of a thermal gradient carries a heat current and an electric current. These two currents represent the linear response of the system to the external fields. The typical setup of a thermopower measurement is the open circuit geometry with $\bvec{j}_{e}=0$, which defines the value of the electric field $\bvec{\mathcal E}$. The Seebeck coefficient, defined in Eq. \eqref{eq:def of Q} as the ratio of the applied thermal gradient and the induced electrochemical potential, is then given by
\begin{align}
     Q\equiv|\bvec{\mathcal E}| / |\bvec{\nabla T}| = K_{11}/K_{12},
\label{eq: Q = K12 / K11}
\end{align}
with the transport coefficients $K_{ij}$ defined by
\begin{align}
    \bvec{j}_{e}=K_{11}\bvec{\mathcal E}+K_{12}\bvec{\nabla T}.
\end{align}

\section{The electrical conductivity}
\label{sec:Numerical simulation of the conductivity as a function of the band filling}

We compute the electrical conductivity $\sigma$ for several temperatures in the range from $k_{B}T=0.01\,t$ to $0.2\,t$ for a generic choice of $t'=0.2\,t$. We scan through the band from very low filling, $\mu\approx -2.8\,t$, to the almost completely filled band, $\mu\approx 4.2\,t$, by tuning the chemical potential (see Fig. \ref{fig: conductivityBand}, right panel). We observe a rich structure in $\sigma(\mu)$, which emphasizes the importance of band-structure effects. In the following we analyze the mechanism leading to these structures for different band fillings.

\subsection{Special band fillings}
\label{subsec:SpecialBandFillings}

We start our discussion of the band filling dependence of the conductivity at low fillings which corresponds to a small, almost circular Fermi surface closed around the $\Gamma$ point. In the previous section we have explained that umklapp scattering is suppressed for sufficiently small Fermi surfaces, i.e., $k_{F} < \pi/(2a)$. To be precise, the absence of umklapp scattering is only true at zero temperature. At finite temperatures umklapp-scattering processes are possible through thermal activation because there are always states with non-vanishing occupation numbers and $k \geq \pi/(2a)$. Thus, the onset of umklapp scattering in the zero-temperature limit is bound to a chemical potential $\mu=\mu_{U1}$ for which the condition $k_{F}=\pi/(2a)$ is first satisfied somewhere on the Fermi surface (cf. Fig. \ref{fig: conductivityBand}, marked by a red circle). A possible umklapp-scattering event could be of the form
\begin{align}
    \{(u,0),(u,0)\}\longmapsto \{\underbrace{(3u,0)}_{(-u,0)},(-u,0)\},
\end{align}
with $u=\pi/(2a)$ and obviously transfers a momentum of $(4u,0)$ to the lattice.

There is also a second kind of umklapp scattering with a momentum transfer of $(4u,4u)$ to the lattice, for which one can find with similar arguments that the Fermi surface must intersect with the so-called umklapp-surface, a diamond that connects the four saddle points of the dispersion at $(2u,0)$ and $(0,\pm 2u)$. Depending on the sign of $t'$, this intersection takes place first at $(\pm u,\pm u)$ for $t'<0$, or at the saddle points for positive $t'$. 
In the limit of zero temperature this umklapp channel is only available if the Fermi surface intersects the umklapp surface and is turned off otherwise. 

When the Fermi surface approaches the saddle points, one observes another intriguing feature of this model. Due to the vanishing slope of the dispersion, the density of states diverges as a van Hove singularity. This property of the density of states is associated with a Lifshitz transition, as the Fermi surface changes from electron like, i.e., closed around the $\Gamma$ point, to hole like, closed around the corner of the Brillouin zone. Due to the flat dispersion in the proximity of the saddle points, the phase space for scattering through thermally activated states is expected to be very large. This large phase space leads to a strong dip in the conductivity (see Fig. \ref{fig: conductivityBand}). The chemical potential at which the Lifshitz transition is located is denoted as $\mu_{L}$. 

Because $\mu_{L}$ also defines the onset of the $(4u,4u)$-umklapp scattering, the end of this umklapp regime is labeled $\mu_{U2} > \mu_{L} $. For $\mu>\mu_{U2}$, the Fermi surface no longer intersects with the umklapp surface. A further umklapp boundary towards the top of the band is again connected to the $(4u,0)$-momentum transfer to the lattice. The corresponding critical chemical potential is denoted as $\mu_{U3}$ and it is related to $\mu_{U1}$ by particle-hole symmetry. We refer to the filling levels that represent boundaries for umklapp scattering as umklapp edges.

In summary, there are four special filling levels in our model, the umklapp edges at $\mu_{U1},\mu_{U2},\mu_{U3}$ and the filling level of the Lifshitz transition, $\mu_{L}$. They give a tag to each anomaly observed in the resistivity or conductivity as a function of the chemical potential $ \mu $. In the absence of umklapp scattering, $ \mu < \mu_{U1} $ and $ \mu > \mu_{U3} $ only impurity scattering constrains the electric transport which, thus, is roughly temperature independent. Moreover, we would like to emphasize again the pronounced dip in the conductivity is associated with van Hove filling.

\begin{figure*}[t]
\begin{center}

\includegraphics[width=0.49\textwidth]{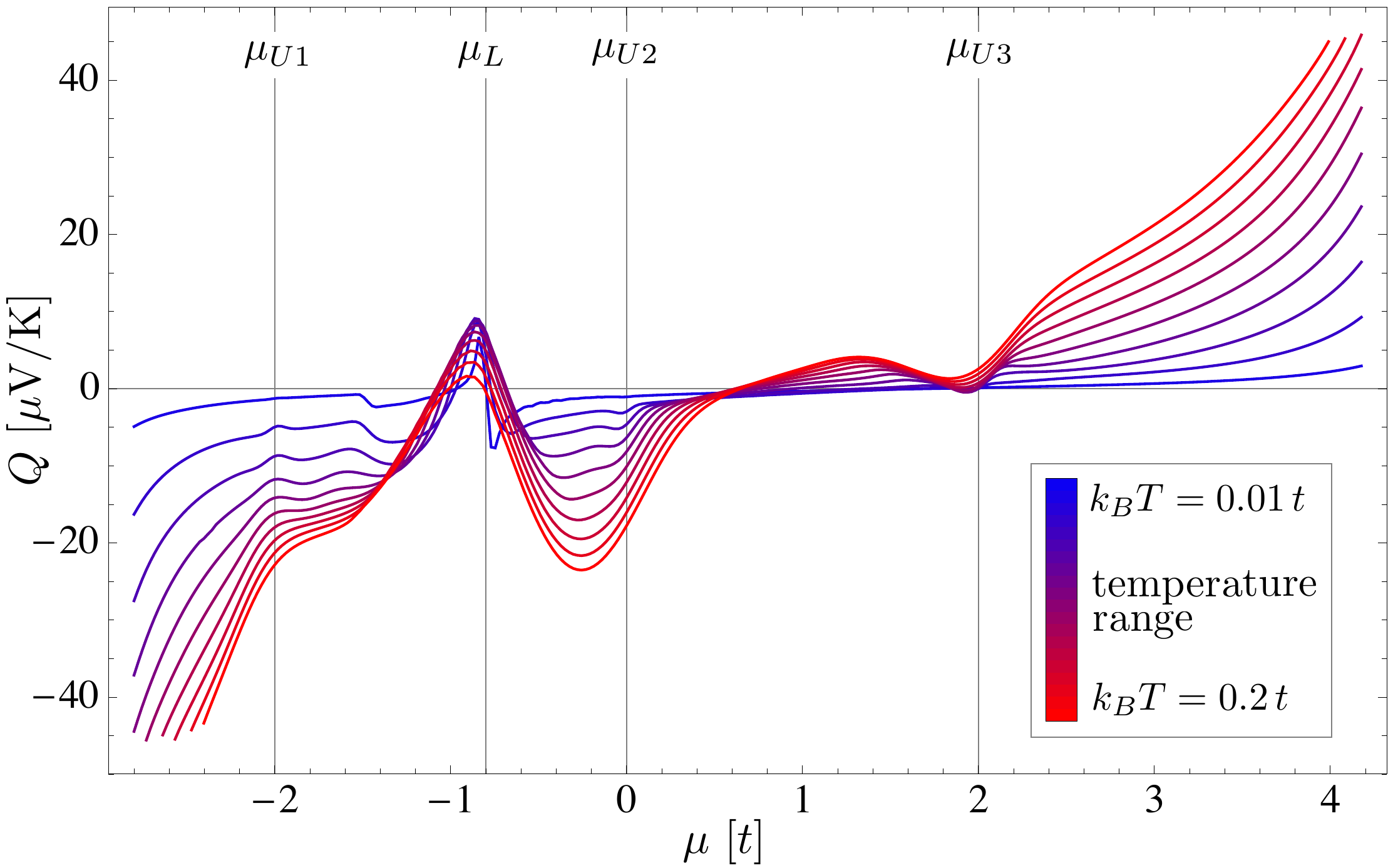}
\includegraphics[width=0.49\textwidth]{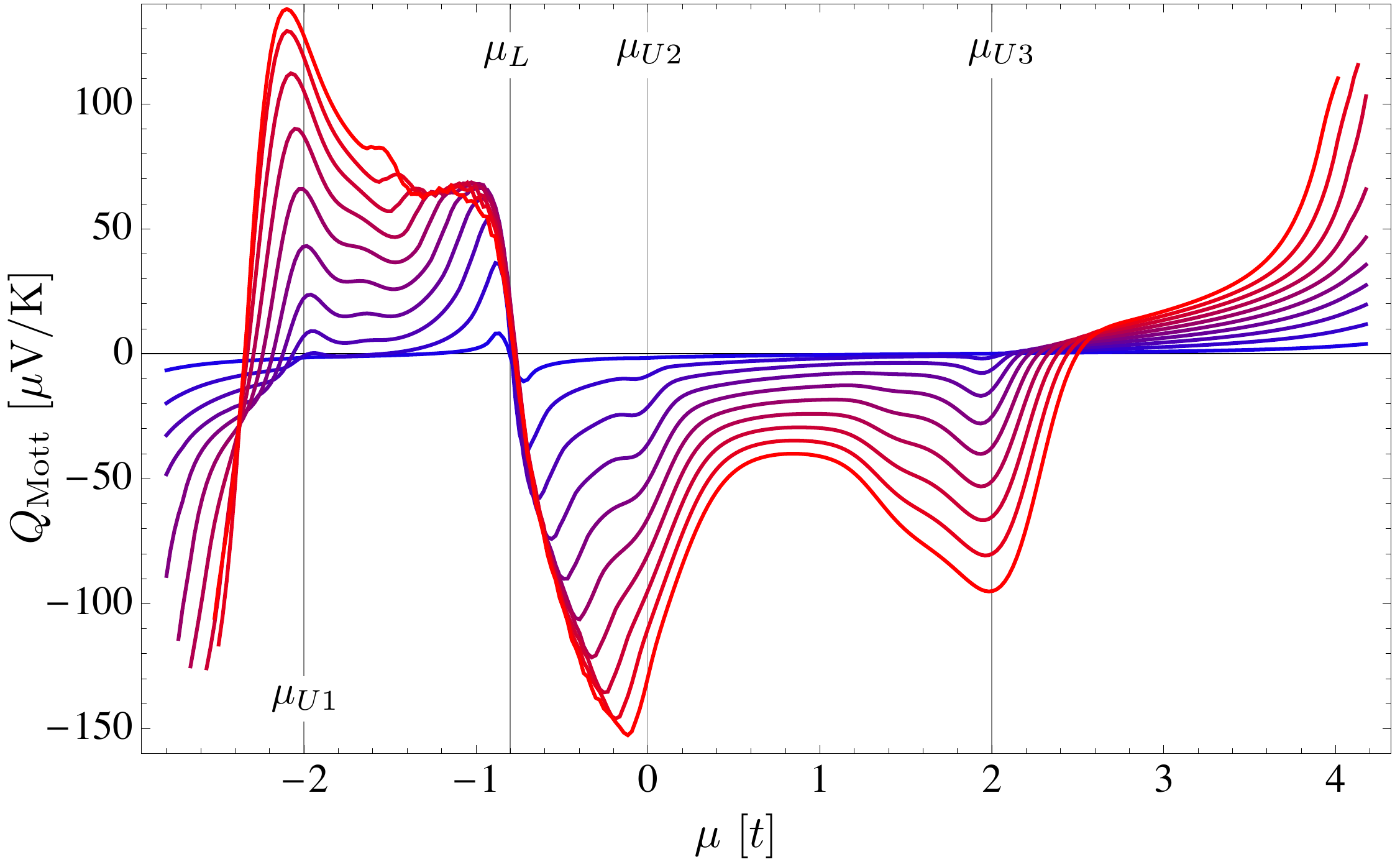}
\caption{\emph{left}: Numerical results for the Seebeck coefficient $Q=K_{12}/K_{11}$ [cf. Eq. \eqref{eq: Q = K12 / K11}] as a function of the band filling with the same dispersion as for the conductivity. \emph{right}: Calculation of $Q_{\text{Mott}}$ within Mott's formula [cf. Eq. \eqref{eq: Mott}]. Note that Mott's formula yields especially bad predictions at the umklapp edges $\mu_{U1},\mu_{U2},\mu_{U3}$.}
\label{fig: seebeckBand}

\end{center}
\end{figure*}

\section{Thermopower and the limitations of Mott's formula}
\label{sec:Thermopower and the breakdown of Motts formula}

We use now our numerical simulation scheme and compute the Seebeck coefficient $Q$ according to Eq. \eqref{eq: Q = K12 / K11} (Fig. \ref{fig: seebeckBand}, left panel) and compare it to $Q_{\text{Mott}}$, the prediction of Mott's formula \eqref{eq: Mott} (Fig. \ref{fig: seebeckBand}, right panel).
The differences between $Q$ and $Q_{\text{Mott}}$ are striking. Most prominently, at the umklapp edges Mott's formula predicts a largely enhanced Seebeck coefficient, as expected from the strong energy dependence of the conductivity due to the onset of umklapp scattering. Our calculation, however, shows that $Q$ is not enhanced at all, and at $\mu=2\,t$, $Q$ is even strongly suppressed. Moreover, a sign change at the Lifshitz transition point, visible for all temperatures in $Q_{\text{Mott}}$, cannot be directly associated with the Lifshitz transition, at least when two-particle scattering becomes relevant above intermediate temperature, as shown in the plot of $ Q $. Instead, there is a sequence of sign changes away from the Lifshitz point. 

The discrepancy can be understood by scrutinizing the approximations made in the derivation of Eq. \eqref{eq: Mott} \cite{ashcroftmermin}. 
While the Sommerfeld expansion should, in principle, be applicable for $T\ll T_{F}$, Mott's formula rests on the further assumption that scattering rates are essentially isotropic over the whole Fermi surface and are only energy dependent. Obviously this latter condition is not satisfied in the case of umklapp scattering,
which provides strongly anisotropic scattering rates due to strict momentum space constraints. Especially close to the umklapp edges, only small parts of the Fermi surface 
allow for umklapp processes and, therefore, the scattering rates for these states are much greater than for the states unaffected by umklapp scattering.

\begin{figure}[t]
\begin{center}

\includegraphics[width=0.49\textwidth]{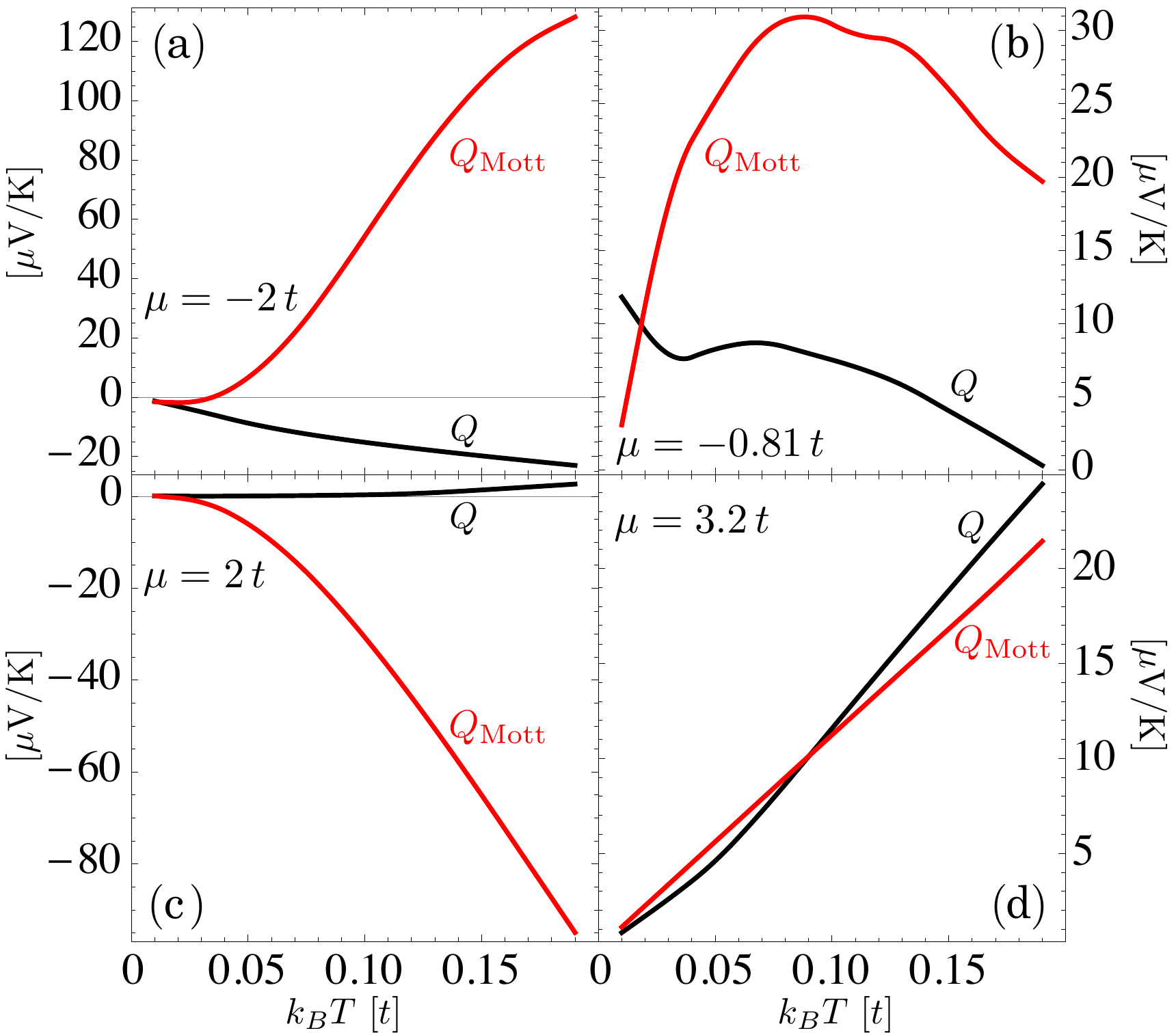}
\caption{Direct comparison of the Seebeck coefficient $Q$ (black lines) with the approximation $Q_{\text{Mott}}$ based on Mott's formula, Eq. \eqref{eq: Mott}.}
\label{fig: seebeckRuns}

\end{center}
\end{figure}

The sign of the Seebeck coefficient is often considered as a way to identify the charge of the quasiparticles. Thus, a sign change is expected at the Lifshitz transition as seen in $ Q_{\text{Mott}}$ \cite{abrokosovMetals}, which represents a transition from an electron like system to a hole like system due to the change of Fermi-surface topology. In fact, sign changes do occur, 
in particular, at low temperatures. The situation looks, however, more complex, as actually several sign changes occur. Counter-intuitively, the sign change at the 
Lifshitz transition (positive to negative) is opposite to what is naively expected, when turning from an electron like (negative $Q$ expected) to a hole like Fermi surface (positive $Q$ expected). This behavior illustrates that the simple interpretation of the sign of the Seebeck coefficient in terms of quasiparticles as particles or holes is not valid for nontrivial situations as encountered, for instance, in lattice systems at intermediate filling levels. In the vicinity of a van Hove singularity, the conductivity shows a strong dip, especially pronounced at low temperature. Due to this dip, the slope of the conductivity as a function of the band filling qualitatively corresponds to $-\sigma'(\mu)$ at the band edge with the same Fermi-surface topology. Thus, Mott's formula [$Q_{\text{Mott}}\sim\sigma'(\mu)]$ predicts a sign change opposite to the naive expectation based on the band edges. 

A  further intriguing feature is the sign change at intermediate temperatures away from the Lifshitz transition, which is 
in contrast to the prediction from Mott's formula. This is also noteworthy, as sign changes are commonly understood as a precise way to detect 
the location of the Lifshitz transition \cite{abrokosovMetals}.

In Fig. \ref{fig: seebeckRuns} we highlight the discrepancy between $Q$ and $Q_{\text{Mott}}$. We plot both $Q$ and $Q_{\text{Mott}}$ as a function of the temperature for four different band fillings. We select filling levels at the first [Fig. \ref{fig: seebeckRuns}(a)] and the third umklapp edge [Fig. \ref{fig: seebeckRuns}(c)] because the scattering rate anisotropy is particularly pronounced due to the restriction of umklapp scattering to only small portions of the Fermi surface. In Fig. \ref{fig: seebeckRuns} (b) we plot the comparison for van Hove filling, while in Fig. \ref{fig: seebeckRuns}(d) we demonstrate that both approaches yield quantitatively comparable results outside of the umklapp regime, where umklapp scattering cannot yield anisotropic scattering rates.

\section{Unconventional thermopower related to the Lifshitz transition and the umklapp edges}
\label{sec:Unconventional thermopower related to the Lifshitz transition and the umklapp edges}

A linear low-temperature dependence of $ Q $, much as the $ T^2 $ behavior of the resistivity, can be interpreted as a hallmark of Fermi-liquid behavior. 
Deviations from this behavior are consequently considered as a signature of unusual, so-called non-Fermi-liquid physics. In this spirit, $Q/T$ has been used as an experimental probe to explore non-Fermi-liquid transport properties. In particular, using the temperature dependence of the Seebeck coefficient as a probe for quantum critical behavior was recently put forward by several groups \cite{QCPfromSeebeckMeasurements,laliberte2011,daou2009}.

\begin{figure}
\begin{center}
\includegraphics[width=0.45\textwidth]{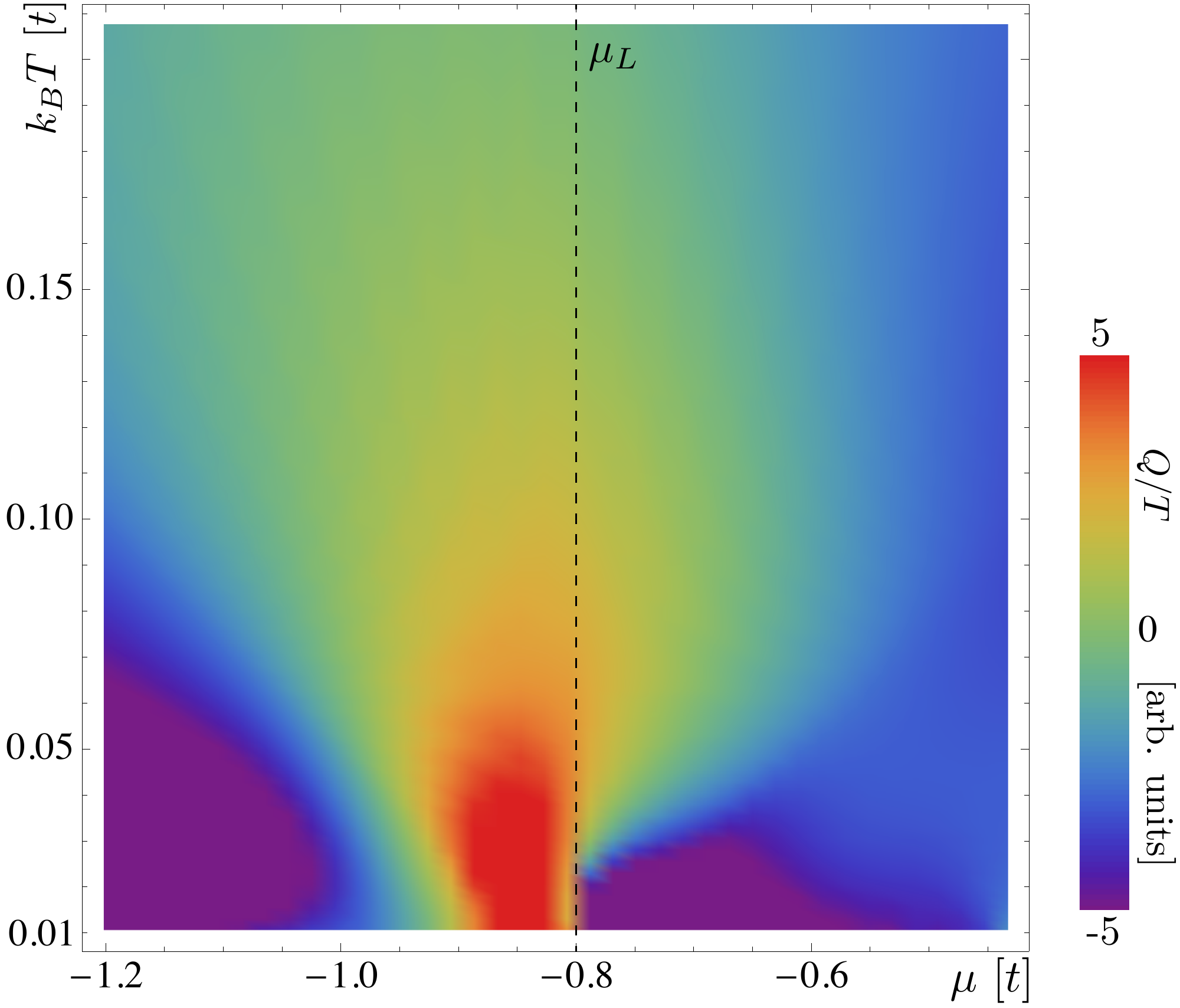}
\caption{Plot of $Q/T$, for which the temperature derivative can be interpreted as a measure of non-Fermi-liquid transport behavior. As a function of the band filling, one finds several sign changes at low temperature. In particular, the low-temperature sign change at the Lifshitz transition $\mu=-0.8\,t$ is pronounced. Note, however, that the sign change is from positive (hole like) to negative $Q$ (electron like). This contradicts the naive expectation that the Seebeck coefficient is directly related to the topology of the Fermi surface.}
\label{fig: seebeckNormalizedL}

\end{center}
\end{figure}

We have investigated the properties of $ Q(T) $ in this respect near the Lifshitz transition as well as near the upper umklapp edge ($\mu_{U3}$). For both cases we observed clear deviations from the expected $Q\sim T$ behavior. 

First, we discuss the thermoelectric effect near the Lifshitz transition which is combined with a singularity in the density of states. In Fig. \ref{fig: seebeckNormalizedL} a phase diagram of $Q/T$ around the Lifshitz transition ($\mu=-0.8\,t$) is shown for the same model as displayed in Fig. \ref{fig: seebeckBand}. We find the multiple sign changes of $Q$ at low temperature and verify that the sign change is opposite to the naive expectation that predicts a change from an electron like Seebeck coefficient to a hole like $Q$, crossing the Lifshitz transition from below. Moreover, $Q/T$ increases towards low temperature, again not in agreement with $Q\sim T$. All the temperature dependencies that can be found in Fig. \ref{fig: seebeckNormalizedL} are evidence for unconventional behavior, however, the deviation from $Q\sim T$ can be even better visualized by taking the temperature derivative of $Q/T$.

\begin{figure}
\begin{center}

\includegraphics[width=0.45\textwidth]{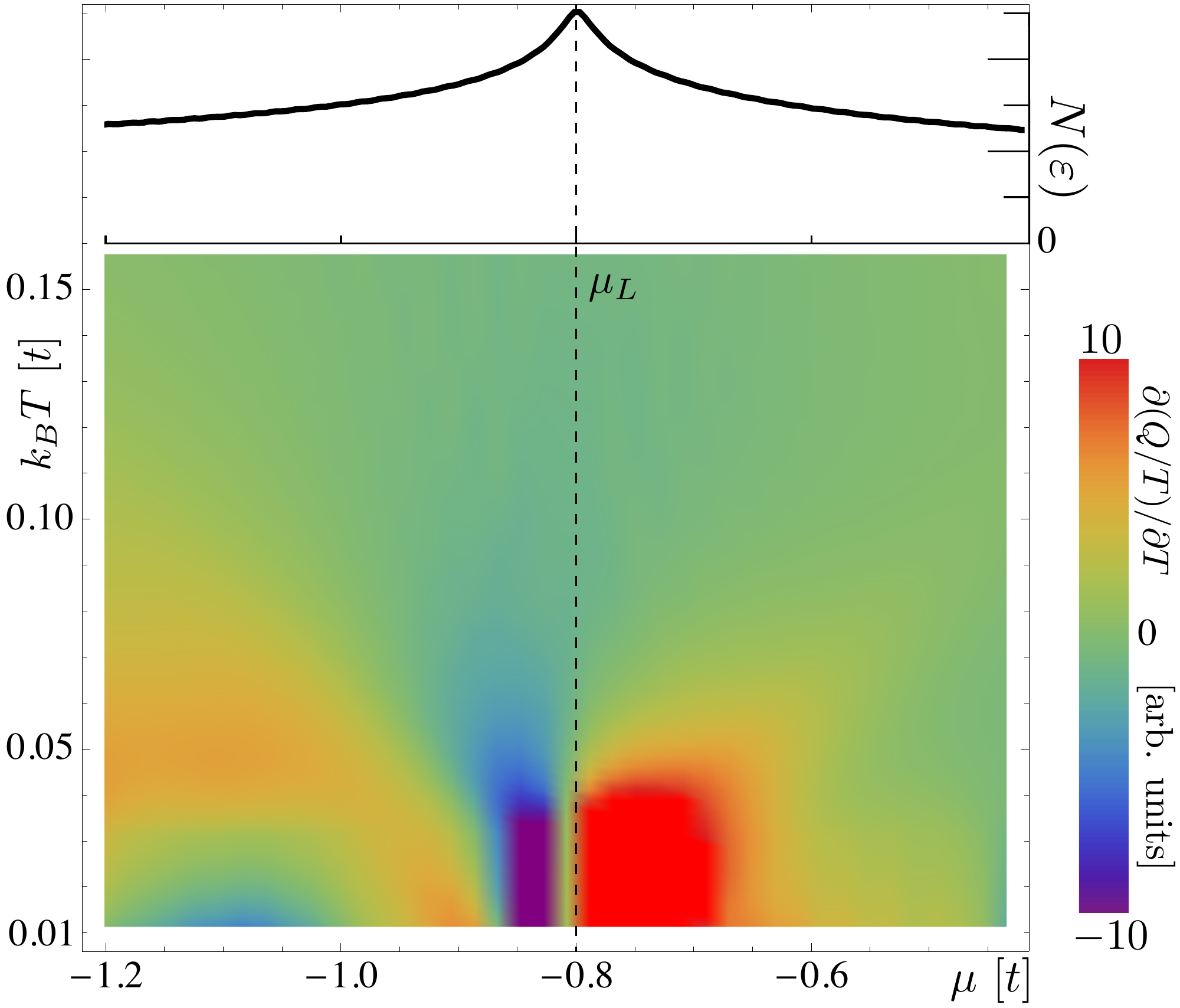}
\caption{Plot of $\partial (Q/T) / \partial T$ in the vicinity of the Lifshitz transition ($\mu_{L}=-0.8\,t$), for which finite values are interpreted as signatures of non-Fermi-liquid transport properties. The color scale was truncated at 0.2 times the maximal value of $\partial (Q/T) / \partial T$. The density of states in this range of chemical potential is shown in arbitrary units.}
\label{fig: seebeckDerivativeL}

\end{center}
\end{figure}

Our quantitative measure for the deviations from the conventional Fermi-liquid behavior, $\partial (Q/T)/\partial T$, is shown in Fig. \ref{fig: seebeckDerivativeL}. For convenience, we have overlaid the phase diagram with a plot of the density of states which shows the presence of the van Hove singularity.

The deviations from a linear-$T$ Seebeck coefficient around the Lifshitz transition are clearly visible from Fig. \ref{fig: seebeckDerivativeL}. The quantity $\partial (Q/T) / \partial T$ shows a pronounced peak at the Lifshitz transition in the low-temperature limit and it has an opposite sign on both sides of the transition, just as the Seebeck coefficient itself.

\begin{figure}
\begin{center}

\includegraphics[width=0.43\textwidth]{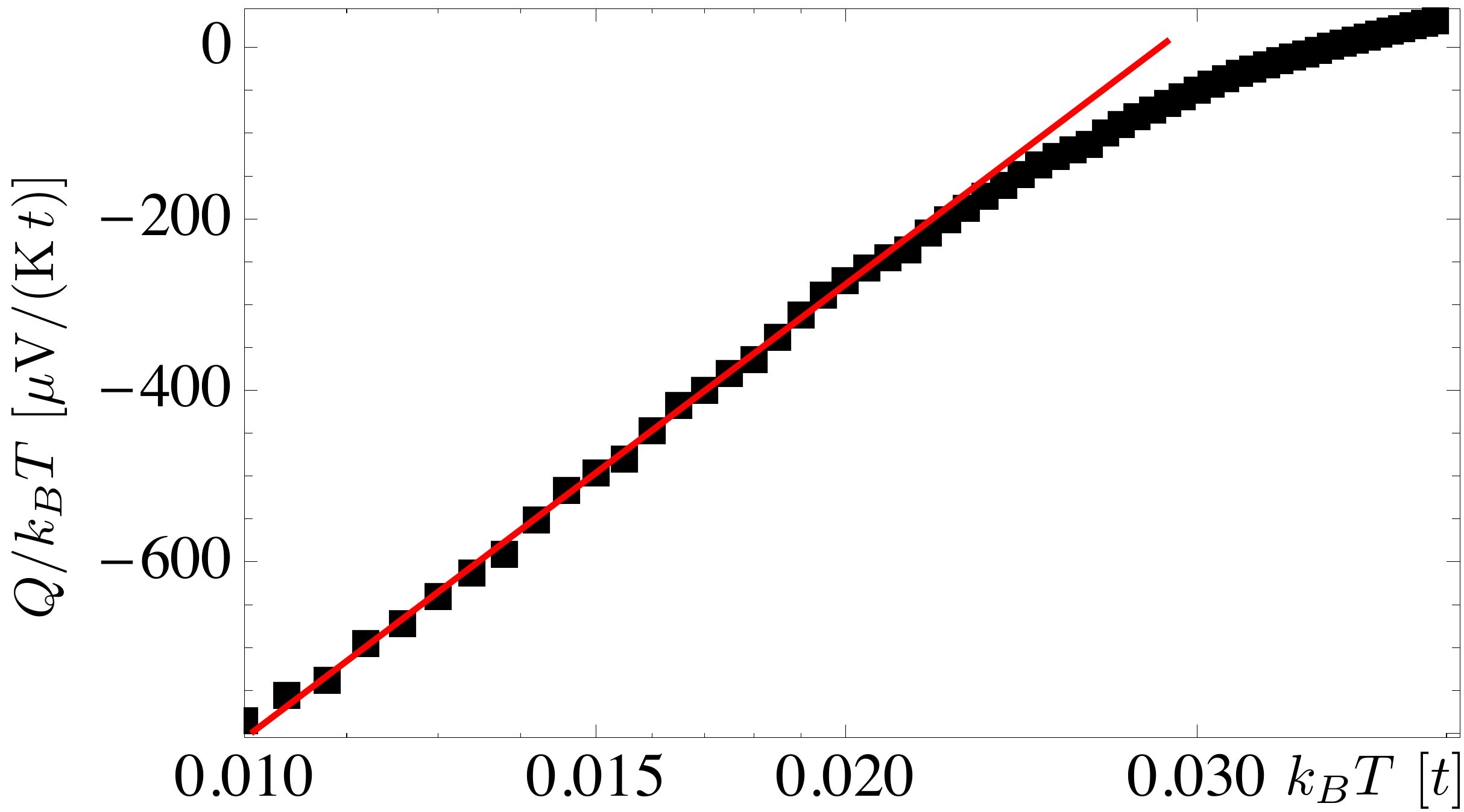}
\caption{Plot of $Q/T$ on a logarithmic temperature scale for band filling which corresponds to $\mu=-0.76\,t$. The low-temperature linearity in this plot is a signature of the $Q\sim \log(T)$ dependence which is often associated with the presence of a QCP.}
\label{fig: Q/T logplot}

\end{center}
\end{figure}

As mentioned above, large deviations from standard Fermi-liquid behavior are often interpreted as evidence for a non-Fermi-liquid state in the vicinity of a QCP. Another property that is considered as a typical feature is a low-temperature logarithmic divergence of $Q/T$\cite{laliberte2011,daou2009}. Motivated by the unconventional properties of $Q/T$ that we have observed so far, we have plotted the low-temperature regime of $Q/T$ on a logarithmic temperature scale for a band filling level close to the Lifshitz transition (Fig. \ref{fig: Q/T logplot}). We clearly identify a regime below $k_{B}T\approx 0.02\,t$ in which $Q/T$ grows logarithmically with $T\rightarrow 0$. This behavior is indicated by the linear fit in the graph. It is intriguing to find signatures that are typical for a QCP also to appear generically in the vicinity of a Lifshitz transition without the loss of quasiparticle properties of the carriers. It is well known that the singular form of the density of states can alter generic scaling properties. 

We now focus on the umklapp edges which are examples for unconventional transport behavior that is not related to any singular feature in the density of states. Unconventional transport properties in this case can be attributed to the strong anisotropy in the scattering rates along the Fermi surface near umklapp edges, as indicated by the strong deviations between $ Q $ and $ Q_{\text{Mott}} $. Let us focus here on the third umklapp edge at $ \mu = \mu_{U3} $ (see Fig. \ref{fig: conductivityBand}).

\begin{figure}
\begin{center}

\includegraphics[width=0.45\textwidth]{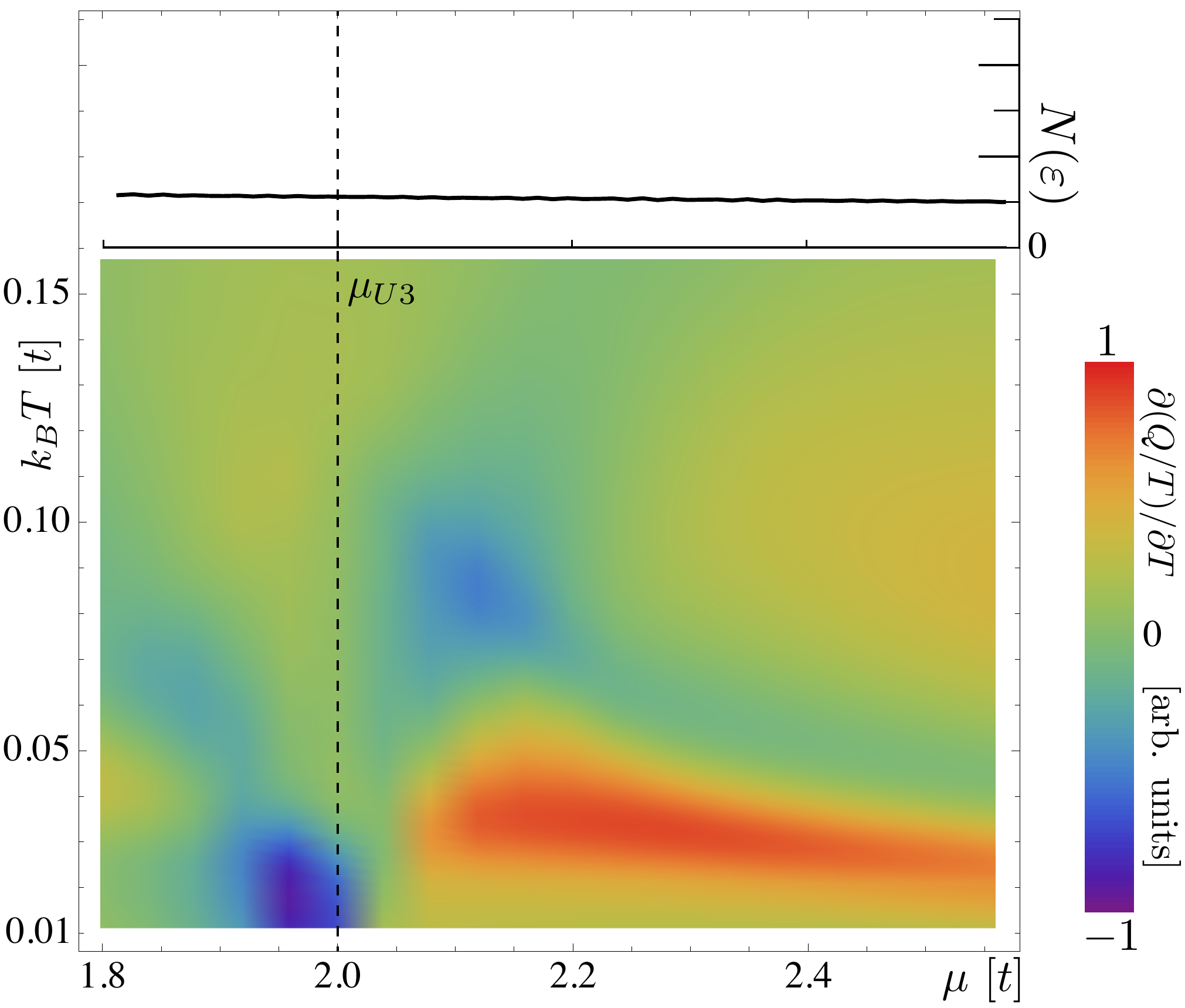}
\caption{Plot of $\partial (Q/T) / \partial T$ in the vicinity of the third umklapp edge ($\mu_{U3}=2\,t$), for which finite values are interpreted as a signature of non-Fermi-liquid transport properties. The color scale covers the entire range from the minimal to the maximal value of $\partial (Q/T) / \partial T$. The density of states in this range of chemical potential is shown in arbitrary units.}
\label{fig: seebeckDerivativeU3}

\end{center}
\end{figure}

In Fig. \ref{fig: seebeckDerivativeU3} we show a plot of $\partial (Q/T) / \partial T$ as we did for a region close to the van Hove singularity at the Lifshitz transition. Note, however, that the maximal value of the color scale is reduced by a factor of 10 compared to the case around the Lifshitz transition shown in Fig. \ref{fig: seebeckDerivativeL}. Nevertheless, we find a complex structure in the temperature versus chemical potential diagram, with multiple sign changes of the quantity $\partial (Q/T) / \partial T$. This rich structure is a consequence of the onset of umklapp scattering for $\mu \leq 2\,t$ in the zero-temperature limit. Moreover, also the temperature driven change of the dominant scattering mechanism from impurity scattering to two-particle collisions has nontrivial effects. The change of the dominant scattering mechanism is associated with a crossover from isotropic to highly anisotropic scattering rates as two-particle scattering is dominated by the Fermi-surface geometry. It is important to note that the deviations from the standard Fermi-liquid results are not related to a feature in the density of states, as is illustrated by the smooth form of $N(\varepsilon)$ in the top of Fig. \ref{fig: seebeckDerivativeU3}.

\begin{figure}
\begin{center}

\includegraphics[width=0.4\textwidth]{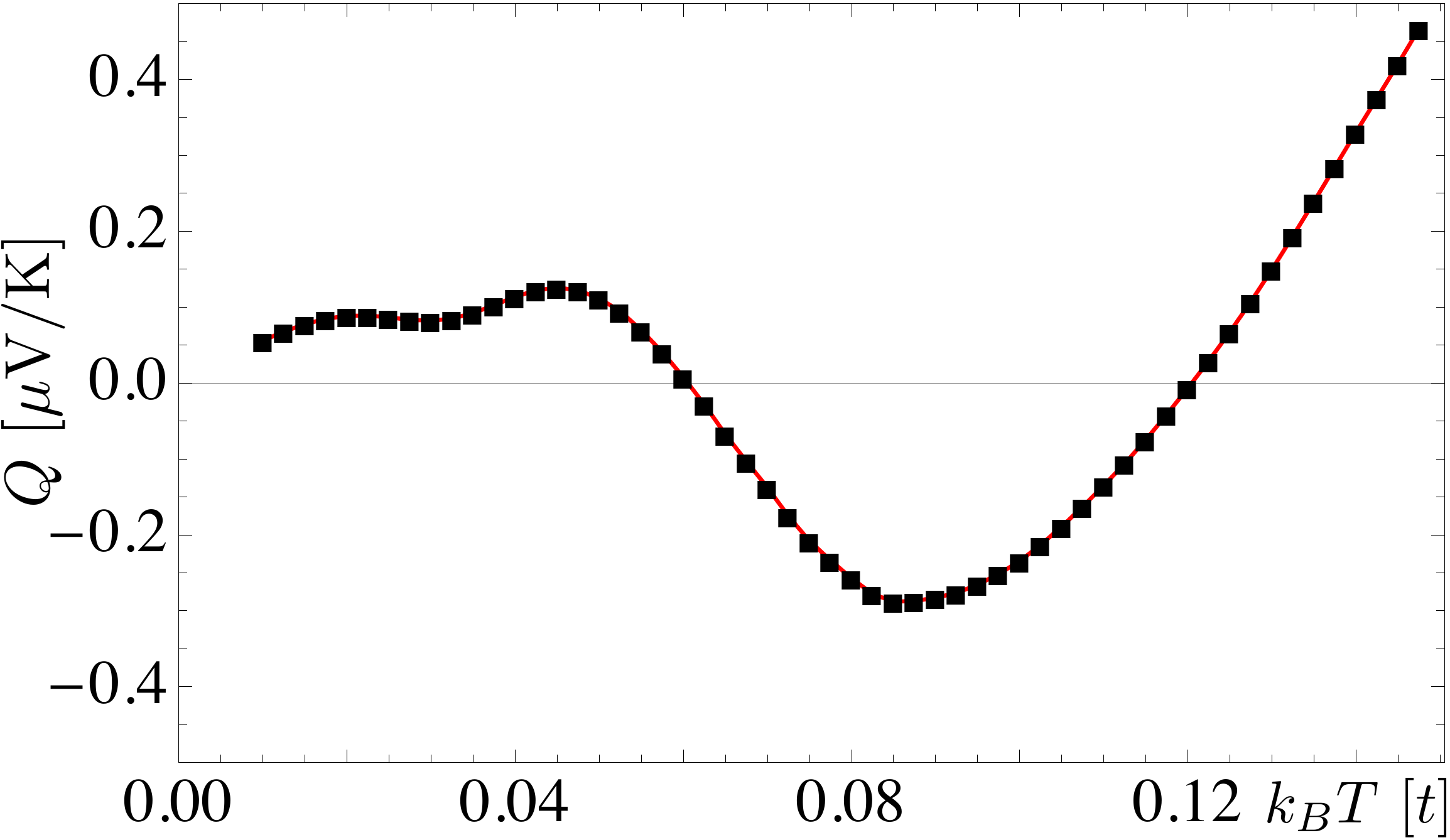}
\caption{Plot of the Seebeck coefficient as a function of the temperature for $\mu=1.88\,t$, close to the third umklapp edge $\mu_{U3}=2\,t$. Note that $Q$ changes sign twice as a function of the temperature.}
\label{fig: seebeckSignChange}

\end{center}
\end{figure}

Figure \ref{fig: seebeckSignChange} illustrates the unusual behavior of $Q(T)$ close to the third umklapp edge. The anisotropy of the scattering rates coefficient results in multiple sign changes of the Seebeck with temperature. Sign changes of the Seebeck coefficient with temperature are also found experimentally as, e.g., in the cuprates \cite{chang_nernst_2010,laliberte2011}. Our simple model shows that such features can appear generically as a consequence of anisotropic scattering, in this case induced by the proximity to an umklapp edge. 

The natural appearance of sign changes of $Q$ with temperature is a further indication that the sign of the Seebeck coefficient cannot be simply interpreted in terms of electron like or hole like quasiparticles. In our model such a simple scheme would only apply in the limit of low carrier concentration or almost full band filling where only isotropic impurity scattering dominates the transport properties.

\section{Discussion and Conclusions}

We have studied the electrical conductivity and the Seebeck coefficient of the thermoelectric effect for an interacting two-dimensional Fermi liquid. We have investigated this model by a numerical solution of the Boltzmann transport equation with a full resolution of angular and radial degrees of freedom in the collision integral. Special geometric constraints for two-particle scattering induce a rich structure in the temperature and band filling dependence of the electrical conductivity and the Seebeck coefficient. Particularly important are the filling levels related to the onset of umklapp scattering and the critical filling level of the Lifshitz transition. While it is widely believed that Mott's formula provides a good description of the Seebeck coefficient for standard metals, our numerical simulation demonstrates that strongly anisotropic scattering rates due to electron-electron scattering that involves umklapp processes spoils this simple approach. While at very low temperatures Mott's formula applies as the effect of electron-electron scattering
is diminished, our analysis of the temperature dependence shows a pronounced quantitative as well as qualitative difference with increasing temperature. Strong anisotropy in the scattering rates is especially pronounced at filling levels close to the umklapp edges. We have demonstrated that the presence of umklapp edges can also lead to deviations from the standard Fermi-liquid behavior expected for $ Q $ at low temperature. 
The van Hove singularity accompanied by a Lifshitz transition in our model constitutes another interesting band filling. The analysis of $Q/T$ shows strong signatures of non-Fermi-liquid behavior for a certain range around the van Hove filling. We also observe here a logarithmic temperature dependence $Q/T\sim \log(T)$ for low temperatures, similar to features usually associated with quantum critical properties.

In this context it is important to note that the density of states does not contain information about the scattering geometry, and thus a calculation based on the structure of the density of states alone cannot yield the features we have found. Figures \ref{fig: seebeckDerivativeL} and \ref{fig: seebeckDerivativeU3} allow for a comparison in this respect. It is important to notice that the feature of the van Hove singularity is small compared to the large background of density of states, and is thus unlikely the leading cause of the strongly anomalous transport behavior. 

We have also discussed an example of a filling level for which the Seebeck coefficient changes its sign as a function of the temperature. This non-trivial temperature dependence is generically found in metals with strong umklapp scattering and reflects the fact that the transport of an electron system on a lattice cannot be uniquely considered to be either particle like or hole like. 

In conclusion, our study demonstrates that even in simple metals many unconventional transport properties are found as a consequence of two-particle collisions that require special Fermi-surface geometries to allow for momentum relaxation. Signatures of critical transport properties, such as $Q/T\sim \log(T)$ or temperature induced sign changes of the Seebeck coefficient, appear naturally due to strong anisotropy in the scattering rates, induced by umklapp scattering. Moreover, the Lifshitz transition shows characteristics resembling those at a quantum critical point, while keeping the quasiparticle nature of the carriers untouched. 
 Further deviations from standard Fermi-liquid physics are associated with the umklapp edges, which go beyond simple effects induced by the structure of the density of states.

\begin{acknowledgements}
	We are grateful to S. Arsenijevi\'{c}, J. Flouquet, L. Forr\'{o}, G. Knebel, D. M\"uller, M. Ossadnik, S. Poppulo, A. Pourret, T. M. Rice, and A. Weidenkaff for helpful discussions. This study was supported by the Swiss Nationalfonds, the NCCR MaNEP, the HITTEC project of the Competence Center Energy \& Mobility, and the Sinergia TEO, as well as by the Toyota Central R \& D Laboratories, Nagakute, Japan.
\end{acknowledgements}

\begin{appendix}

\end{appendix}

\end{document}